# Modified PSO based PID Sliding Mode Control using Improved Reaching Law for Nonlinear systems

Kirtiman Singh, Prabin Kumar Padhy

**Abstract**: In this paper, a new model based nonlinear control technique, called PID (Proportional-Integral-Derivative) type sliding surface based sliding mode control is designed using improved reaching law. To improve the performance of the second order nonlinear differential equations with unknown parameters modified particle swarm intelligent optimization (MPSO) is used for the optimized parameters. This paper throws light on the sliding surface design, on the proposed power rate exponential reaching law, parameters optimization using modified particle swarm optimization and highlights the important features of adding an integral term in the sliding mode such as robustness and higher convergence, through extensive mathematical modeling. Siding mode control law is derived using Lyapunov stability approach and its asymptotic stability is proved mathematically and simulations showing its validity. MPSO PID-type Sliding mode control will stabilize the highly nonlinear systems, will compensate disturbances and uncertainty and reduces tracking errors. Simulations and experimental application is done on the non-linear systems and are presented to make a quantitative comparison.

## I. INTRODUCTION

Nonlinear dynamical systems are considered as an interesting topic of investigation as all real systems are always nonlinear in nature and approximating it linearly can impose severe restrictions on their working range or may give unfeasible result [1]. Also, nonlinear behavior is sometimes introduced in feedback control as they can provide for even better performance than the linear ones [2]. The control laws governing dynamical system need to have nonlinear and time varying behavior with various uncertainties, disturbances and parasitic dynamics, the influence of which has to be carefully taken into account when considering the performance of the system. The principle objective of control law is to generate a control or regulation system which is robust to external disturbances and model uncertainty. The control system has seen development ranging from the ubiquitous PID controller through to high performance controllers. PID controllers are one of the most popular and simple controller used by the engineers mainly in process industries, automatic control etc. Adding to the value of PID's are optimal and robust controllers [10-16]. These control algorithms are independent of the model, linear in nature and their parameter has to be adjusted to obtain an acceptable response characteristic [3]. Recently, there has been a lot of activity in the design of the so-called intelligent control techniques such as fuzzy logic and neural network [5-9], which rely on learning the input-output behavior of the plant to be controlled. Superseding them are the mature mathematically model based controllers in the area of nonlinear control which yield high performance, near world tests of its classes in the aerospace industry. Inverted pendulum is treated as one of the classical problem of control engineering as it has similarity to many complex control situations like helicopter control, humanoid robots stability, space shuttle launching control and robots [18-19]. Instead of directly applying the intelligent techniques, their role has been reversed to processes which are ill defined, nonlinear, complex, time varying and stochastic [17]. Among the existing methodologies, switching control systems are quite prominent as they are quite simple to implement and provide an optimal solution to some control problem. It has its roots lies in relay and bang-bang control. Switching systems originates interesting mathematical problems since they are characterized by ODE (Ordinary Differential Equations) with discontinuous right hand side [27], i.e by changes in the system dynamics, associated by different sets separating each other by a border and the vector fields across the border are directed toward the border itself. Variable structure control (VSC), first proposed and elaborated in 1960's in Soviet Union [28-29], is inherently a nonlinear control technique offering a great many advantages which cannot be achieved using conventional linear controllers. System dynamics in Variable Structure Control is determined not only by the feedback controller but also by the switching strategy. This high speed switching function has its input as the present state of the system and an output as the particular feedback controller produced at that instant, to drive the nonlinear plant's trajectory to a user defined surface. This restriction to the surface represents the controlled system behavior. VSCS enables finite time error convergence. VSCS is a combination of subsystems where each subsystems has a particular control structure and is valid for a specified region, attaining the conventional goals of control as tracking, regulation, stabilization etc.

Prominent characteristics are invariance, robustness and order reduction [30-33], making it a robust or invariant control system. 'Invariance' in loose terms means the system is completely insensitive to parametric uncertainty and external disturbance. Therefore, a principle operation mode of VSC arises, termed as 'sliding mode' in which the border between the sets, defined as the sliding surface encompasses different vector fields directed towards the sliding surface. Sliding surface is an invariant set and under proper conditions the state trajectory becomes independent from the original system dynamics [32-33]. Sliding mode control (SMC) is a nonlinear control method which is simple and easy to implement. The control law with sliding surface is designed such that states move to the sliding surface in a finite time and stays there forever [34-44]. The sliding mode controller is capable of controlling the nonlinear systems over the desired response behavior. Main advantages are the tailored dynamic behavior of the system for a particular choice of switching function and the closed loop response becomes totally insensitive to particular classes of uncertainty and disturbances, and a very quick response time. To specify the performance directly, also makes it an attractive choice for design. The method has gained a lot of attention since early sixties in the former USSR and in the modern world since early eighties (Emelyaunov 60's, Utkin (77, 91), and Levant 87). Sliding mode design approach consists of designing a switching function so that the sliding motion satisfies the design specifications and selection of a control law making the switching function attractive to the boundary manifold [30-35]. Some advantages of the Sliding mode control method are robustness to parameter uncertainty, finite time convergence, insensitivity to bounded disturbances, reduced order compensated dynamics, very fast dynamics response, easy implementation and computational simplicity as opposed to the other robust control approaches [39-41]. After designing the switching surface, next step is to guarantee the existence of a sliding mode. The existence condition for sliding mode is if the velocity vectors of the trajectory always point towards the switching surface, in the vicinity of the sliding manifold $s = 0$. When the trajectory intersects the sliding manifold, the value of the state trajectory remains within a neighborhood encapsulated in a boundary (Slotine and Li, 1991). This existence requires stability of the trajectory on the neighborhood of the sliding manifold i.e. the trajectory must approach the surface at least asymptotically [45]. The motion along the sliding surface $s = 0$ is represented by the phase trajectories and this phase trajectory consists of two modes of the system i.e. reaching phase and sliding phase. While the former denotes the mode in which trajectory, starting from anywhere on the phase plane moves towards the switching manifold and reaches it in finite time and the latter denotes the sliding phase in which the trajectory asymptotically tends to the origin in the region of attraction. Note that the system's robustness is not guaranteed until the sliding mode is reached [43-44]. Designing a sliding mode controller is done by the introduction of a sliding surface so that the tracking error and output deviations converges to zero in finite time. Sliding manifold $s = 0$ represents the transient response of the system during sliding mode and the system dynamics is totally governed by the parameters denoting '$s$'. Sliding mode trajectory also requires an infinitely fast switching mechanism to guarantee the desired sliding behavior. However, Traditional first order sliding mode control has some disadvantages such as chattering effect, limited flexibility with the surface function and constant gain as the error variable. Most fatal one for practical implementation in engineering field is the "chattering" phenomenon, caused due to the 'sign' term in the switching function and it causes the control input to start oscillating around the sliding surface, because of the infinite switching resulting in an unwanted wear and tear. Chattering is caused by switching nonidealities, which exist in the implementation of switching devices and by parasitic dynamics i.e. unmodelled or fast dynamics. If switching gain is large, chattering may cause unpredictable instability. To solve this problem, various approaches have been proposed. Firstly, by using continuous approximations of the discontinuous sign function by smoothening the switching term as the sliding surface reaches zero [58]. Secondly, by disturbance observers and higher order sliding mode allows the problem of chattering to reduce by making the plant's output continuous [44,53]. Lastly sliding sector method forces the state inside the sector, following which Lyapunov function decreases with a specified velocity and zero input [59].

This paper has its main objective on the new sliding surface design and a new power rate exponential reaching law based control design for the tracking and stabilization of the nonlinear systems under uncertainties. The objective is to force the system to attain the sliding surface in finite time and then maintain the system to slide onto the new sliding surface. It results in a system totally insensitive to parameter uncertainties and external disturbances. Firstly PID sliding surface based Sliding Mode Control scheme based on modified particle swarm optimization (MPSO) algorithm is proposed. Secondly a new power rate exponential reaching law has been proposed for faster convergence. Second or direct Lyapunov stability criterion is used to address the issue of stability [47-48]. By MPSO algorithm, the parameters of switching function and PID parameters are optimized to improve the convergence and robustness [20]. A modification in PSO algorithm is done by Chen et.al. 2009 [21].

This proposed algorithm is then implemented on the nonlinear dynamical systems. Firstly, it is implemented in an Inverted Pendulum system. Inverted pendulum is one of the most studied problems in control engineering as well as one of the most complicated dynamical systems [18-19]. Non-Linear inverted pendulum is considered in this paper. Essentially, the system consists of a cart and a pendulum attached to it (Figure 1). The pendulum is balanced upwards by applying a force to the cart thus moving it forward or backward to attain the equilibrium position. Modeling the dynamics of the inverted pendulum yields a highly nonlinear problem. Because of it, system imprecision may come into effect from actual uncertainty in the plant or from a simplified mathematical representation of the plant model dynamics. The aim is to swing-up and stabilize the pendulum from the lower equilibrium position to its upright equilibrium position. Secondly conical tank model is taken into consideration which is a nonlinear TITO system. Controlling the conical tank is a challenging problem because of its nonlinearity and constantly changing area of cross section [77-79]. These systems find their applications in process industries extensively. Its shape, besides giving it a nonlinear nature, is ideal for better drainage in applications. Thirdly van-der pol's is considered as a classical nonlinear equation to prove the proposed algorithm [76]. The paper has been directed towards tracking and stabilization of all the nonlinear systems using the proposed MPSO PID type sliding surface based sliding mode control technique.

This paper has been organized as follows: Section 2 introduces about PID-type sliding surface based sliding mode control. Section 3 describes about modified particle swarm optimization (MPSO). Section 4 details the control law design, with the stability proved in section 5. Section 6 describes the mathematical model of all three nonlinear systems i.e. Inverted Pendulum, Conical tank and van-der pol's equation with simulation results, with conclusion in Section 7.

## II. PID SLIDING MODE CONTROL

Variable structure control law is discontinuous in nature as it switches from on continuous structure to another respective to the evolution of the state vector, confining the system dynamics according to the applied feedback controller as well as to the switching strategy. This multiple control structure acts like a parallel connection of several different continuous subsystems, acting one at a time. By considering the nonlinear dynamic system

$$\dot{x} = f(x,t,u) + g(x,t).u(t) \qquad x \in \mathcal{R}^n, \ u \in \mathcal{R}^m. \tag{1}$$

subjected to discontinuous feedback

$$u_i = \begin{cases} u_i + (x,t) & if \ s_i(x,t) > 0 \\ u_i - (x,t) & if \ s_i(x,t) < 0 \end{cases} \quad (2)$$

Sliding mode control is a fundamental property of Variable structure system (VSS) as it exhibit a particular behavior characterized by the commutation between various system structures taking place at infinite frequency. Sliding mode control alters the dynamics of a nonlinear system by applying a multiple feedback control structure acting on the opposite sides of a predetermined surface in the system state space. Each of the controllers is designed so as to ensure that the representative trajectory is pushed towards the surface, so that it approaches the surface and once it hits the surface for the first time, it stays on it afterwards. The resulting motion of the system is restricted to the surface which is interpreted as 'sliding' of the system along the surface. This motion of the system as it slides along the manifold is called sliding mode and the geometrical locus is called sliding surface. Sliding surface is defined as

$$S(x,t) = (\frac{d}{dt} + \lambda)^{(n-1)}.e \quad (3)$$

Sliding surface is described by $s = 0$ and the sliding mode commences after a finite time when the system trajectories reached the surface. Sliding mode is characterized by a very high robustness feature and is insensitive to model uncertainties and disturbances. A single input nonlinear systems second order equation is represented as:

$$\ddot{\theta} = f(\theta, \dot{\theta}, t) + g(\theta, t).u(t) + d(t) \quad (4)$$

where, $f(\theta, \dot{\theta}, t) \ and \ g(\dot{\theta}, t)$ are known nonlinear functions, embodying plant parameters, model uncertainties and unmodelled dynamics.

$d(t)$ =bounded external disturbances.

t = independent time variable.

$\theta$ =output at any particular time, $\theta(t) \epsilon \mathcal{R}$.

$u(t)$= control action generated.

The control problem is to asymptotically track the desired response using a discontinuous control law generated by the sliding mode algorithm, in the presence of parameter variations, model uncertainties and bounded external disturbances. To track the desired trajectory, error between the desired and the reference trajectory is defined as:

$$e(t) = \theta_r(t) - \theta(t) \tag{5}$$

In the first order sliding mode controller design, sliding surface is chosen with a relative degree of one w.r.t. the control input, i.e. control input acts on the first derivative of the sliding surface. Convergence entails the problem of stability of the origin in an m-dimensional manifold, therefore existence condition has to be formulated in terms of generalized stability theory, using the second method of Lyapunov for analysis. This requires selecting a generalized Lyapunov function $V(t,x)$ which should be positive definite and should have negative time derivative in the region of attraction [66-69].

To generate a control for the PID-type sliding surface, which has Proportional Integral and derivative error term will remove the traditional Sliding Mode Controller deficiency. PID sliding surface is defined as:

$$s(t) = K_p e(t) + K_d \dot{e}(t) + K_i \int_o^t e(\mathcal{T}) d\mathcal{T} \tag{6}$$

where $K_p, K_i, K_d$ are positive coefficients providing flexibility for sliding surface design and $K_p, K_i, K_d \in \mathcal{R}^+$ and $s(t) \in \mathcal{R}$. Assuming that the system is initiated at the region $s(t) > 0$, will result in increasing $s(t)$ and an insufficient input to drive the error to converge to the sliding manifold. Integral action increases the control action to force the error to the siding manifold leveling with the reaching condition $\dot{V}(t) < 0$. Control action is reduced as $s(t) \to 0$ forcing the system into the sliding mode. This problem of forcing to the sliding surface, constraining the error towards zero is equivalent to that of set point tracking for $t > 0$. Proportional and derivative action acts on the stabilization part of the control action. The control forces for the early convergence of the error to the sliding manifold, which in turn amplifies the integral step leading to a better switching manifold convergence.

A first order sliding mode control is defined if and only if $s(t) = 0$ and fundamental criterion as $s(t) \cdot \dot{s}(t) < 0$. Aim of the controller, is to force the tracking error $e(t)$ to converge on the sliding surface $s(t) = 0$. By maintaining the error e(t) onto the sliding surface $s(t)$ will lead to e(t) approaching zero as $t \to \infty$. After reaching the sliding surface s(t), error will simply slide along the sliding hyper-surface, into the origin e(t)=0 even in the presence of matched parameter uncertainties, model uncertainties and bounded disturbances. The error after starting with an initial value will converge to the boundary layer and slide along it until it reaches $\dot{e}(t) = 0$.

First order derivative of the sliding manifold (PID surface) is represented as

$$\dot{s}(t) = K_i e(t) + K_p \dot{e}(t) + K_d \ddot{e}(t) \tag{7}$$

On deriving the error to second order, we get

$$\dot{e}(t) = \dot{\theta}_r(t) - \dot{\theta}(t) \tag{8}$$

$$\ddot{e}(t) = \ddot{\theta}_r(t) - \ddot{\theta}(t) \tag{9}$$

Substituting the error equation of $\ddot{\theta}(t)$ into the second derivative and writing the resultant first derivative surface equation, we get

$$\dot{s}(t) = K_i\big(\theta_r(t) - \theta(t)\big) + K_p\big(\dot{\theta}_r(t) - \dot{\theta}(t)\big) + K_d\big(\ddot{\theta}_r(t) - \ddot{\theta}(t)\big) \tag{10}$$

Control input is defined as

$$u(t) = u_{eq} + u_{sw} \tag{11}$$

where, $u_{eq}$ = Equivalent / Reaching control, and

$u_{sw}$ = Switched control.

## III. MODIFIED PARTICLE SWARM OPTIMIZATION

Particle swarm optimization (PSO) is the evolutionary computation technique proposed by Kennedy and Eberhart in 1995. Evolutionary algorithms are generic population based search approach, inspired by biological evolution i.e. reproduction, mutation, recombination and selection. Evolutionary Algorithms (EAs) are more elegant than traditional optimization techniques as they generate a population of potential solution rather than a single point solution. EAs have to proceed through several iterations and after each iterations, solutions having better fitness value are selected over poor solutions and offspring are generated under some rule using these selected solutions.

### A. STANDARD PARTICLE SWARM OPTIMIZATION (PSO)

Particle swarm optimization (PSO), derived from the social-psychological theory, is a type of evolutionary algorithm where each potential solution called 'particle' changes its position within the solution space with time, where each particle is the one-dimensional search space without the weight and size of the particle. During the iterations, each particle adjusts its position according to the best position obtained by the whole group as well as by that particular particle itself, termed as 'Velocity location search model'. The swarm direction of a particle is decided by the set of neighboring particle and the history of its own experience. Let $X_i$ and $V_i$ represent the $i^{th}$ particle position and its corresponding velocity in search space. The best previous position of $i^{th}$ particles recorded and represented as $Pb_i$. The best particle among all the particles in the group is represented as $Gb$. The updated velocity of $i^{th}$ individual is given as

$$V_i^{k+1} = w^k V_i^k + c_1 r_1 (Pb_i^k - X_i^k) + c_2 r_2 (Gb^k - X_i^k) \qquad (12)$$

and the next generation value as follows

$$X_i^{k+1} = X_i^k + V_i^{k+1} \qquad (13)$$

where, $V_i^k$ - Velocity of $i$<sup>th</sup> individual at iteration $k$

$w^k$ - Inertia weight at iteration $k$

$c_1 c_2$ - Acceleration factors

$r_1 r_2$ - Uniform random numbers between 0 and 1

$X_i^k$ - Position of $i$<sup>th</sup> individual at iteration $k$

$Pb_i^k$ - Best position of $i$<sup>th</sup> individual at iteration $k$

$Gb^k$ - Best position of the group until iteration $k$

**B. MODIFIED PARTICLE SWARM OPTIMIZATION (MPSO)**

To improve the convergence characteristics of the PSO algorithm, modifications have been implemented in it. Evolution equation of Modified PSO, proposed by Chen et. al., has been modified to ensure that the proposed algorithm has strong global convergence capability initially while it has strong local convergence capability later. In the Modified PSO algorithm $C_1$ decreases exponentially and $C_2$ increases exponentially with time. Effect of taking time dependent values for $C_1$ and $C_2$:

1. The dependence of particle's next position on its previous best position decrease with time.
2. The dependence of particle's next position on global best position increases with time.
3. The main strength of PSO is its property of having random particles but after sometime when the swarm has met some potential solutions and one of them is best possible till then. After this time, the swarm should search around that best solution only, because it is highly likely that the optimum value for fitness function is lying around near to that best solution.
4. It is highly unlikely that particles scattered around the full solution space will give better results than particles concentrated around a smaller subspace which contains the solution.
5. The whole swarm concentrates around the single global best point in very little time and continues their search for global optimum in a very small subspace around global best. Swarm concentrating around a small subspace results in better and quicker optimization.

The expressions for updating the velocity and position of each particle are

$$V_i^{(i+1)} = w * V_{in}^{(i)} + C_1 * \left(P_{in} - X_{in}^{(i)}\right) + C_2 * \left(G_{in} - X_{in}^{(i)}\right) \tag{14}$$

$$X_{in}^{(i+1)} = X_{in}^{(i)} + V_{in}^{(i+1)} \tag{15}$$

where,

$$w = 2 - \left(1 + \frac{1}{2k_{max}}\right)^i, \quad C_1 = e^{(-0.05t)} \text{ and } \quad C_2 = \frac{e^{(0.05t)}}{(1 + 0.05e^{(0.05t)})} \tag{16}$$

$i=$ current number particle,

$k_{max} =$ maximum number of iteration.

The objective or fitness function used here is integral squared error, as follows.

$$J(t) = \int_0^T e^2(t) dt \tag{17}$$

MPSO algorithm is used here to choose the parameters of PID-SMC method i.e. the switching and function parameters, which will be working off-line. It is generating a number of solutions for equal number of particles, and its quality will be evaluated by objective function. The objective function will take into account both the speed of reaching manifold and amplitude of chattering.

## IV. CONTROL LAW DESIGN

- **Equivalent control**

First proposed by Utkin, is based on system with estimated plant parameters with zero disturbances. It can be defined as the smooth control law that can be used to determine the system motion restricted to the switching surface $s(x) = 0$, when initial error is precisely located on the sliding surface $s(t)$. For equivalent control (10) is written as $\dot{s}(t) = 0 \text{ with } d(t, u(t)) = 0$.

$$\dot{s}(t) = K_i\big(\theta_r(t) - \theta(t)\big) + K_p\left(\dot{\theta}_r(t) - \dot{\theta}(t)\right) + K_d\left(\ddot{\theta}_r(t) - \ddot{\theta}(t)\right) = 0 \tag{18}$$

Substituting from (4),

$$K_i\big(\theta_r(t) - \theta(t)\big) + K_p\left(\dot{\theta}_r(t) - \dot{\theta}(t)\right) + K_d\left(\ddot{\theta}_r(t) - f(\theta, \dot{\theta}, t) - g(\theta, t).u(t)\right) = 0 \tag{19}$$

Therefore, Equivalent control is given as

$$u_{eq} = -\frac{1}{K_d g(\theta, t)}\left(K_i\big(\theta_r(t) - \theta(t)\big) + K_p\left(\dot{\theta}_r(t) - \dot{\theta}(t)\right) + K_d\left(\ddot{\theta}_r(t) - f(\theta, \dot{\theta}, t)\right)\right) \tag{20}$$

where, $\theta(t) =$ constant according to the requirement.

$K_p, K_i, K_d$ have to selected properly so that the tracking error converge to zero, constraining to the

sliding surface leading to a condition where the polynomial $(K_i e(t) + K_p \dot{e}(t) + K_d \ddot{e}(t) = 0)$ should be Hurwitz.

Hurwitz polynomial defines that the roots must lie on the left side of the plane implying $\lim_{t \to \infty} e(t) = 0$ means globally asymptotic stability for the closed loop system: $e(t) \to 0$ as $t \to \infty$.

- **Exponential Power rate reaching law design**

To improve the convergence capability of the equivalent control approach, reaching laws [80] are used defined in [21-26]. Sliding mode control law based on the Reaching law comprises the reaching phase that drives the system to a stable manifold and the sliding phase which ensures system to slide to the equilibrium. Constant rate reaching law and Exponential reaching laws are defined [32] as

$$\dot{s} = -k_1 s - \varepsilon_1 sign(s) \tag{21}$$

The Proposed scheme is termed as 'Power rate exponential reaching law', is defined as:

$$\dot{s} = -ks - k_{sc}|s|^\alpha . sign(s) \tag{22}$$

where $k > 0, k_{sc} > 0, 0 \le \alpha \le 2$ and sat(s) from (27).

By satisfying the reaching condition (26), the phase trajectory of the system can reach the sliding manifold in finite time and stays at that state. In the reaching condition, the reaching speed will become faster as the term become larger, but will also results in more chattering in the neighborhood of the sliding manifold $s = 0$. By decreasing the magnitude of the reaching condition term, finite reaching time will increases due to the reduction in the convergence velocity leading to lower chattering in the neighborhood of the sliding manifold $s = 0$. Results were compared in the Results and Simulations section. Combining (10) and (22) will result in:

$$K_i(\theta_r(t) - \theta(t)) + K_p(\dot{\theta}_r(t) - \dot{\theta}(t)) + K_d(\ddot{\theta}_r(t) - f(\theta, \dot{\theta}, t) - g(\theta, t).u(t)) = -ks -$$

$$k_{sc}|s|^\alpha . sign(s) \tag{23}$$

- **Switching control**

The switching control is the discontinuous control law which forces the system onto the sliding. It allows an easy verification of the sufficiency conditions for the existence and reachability of the sliding mode i.e. $s(x).\dot{s}(x) < 0$ when $s(x) = 0$.

The switching control $u_{sw}(t)$ is defined as (constant gain)

$$u_{sw}(t) = -\frac{k_{sc}}{g(\theta,t)} sgn(s(x)) \quad (24)$$

where $k_{sc}$ is positive constant and $k_{sc} > d_{max}$ dominates matching uncertainties. Therefore,

$$\dot{s}(t) = -k_{sc} sign(s(t)) - d(t) \quad (25)$$

$$sign(s) = \frac{s}{|s|} = \begin{cases} 1, & s > 0 \\ 0, & s = 0 \\ -1, & s < 0 \end{cases} \quad (26)$$

Boundary Layer is the approach used to reduce the effect of chattering by constraining the control, along a boundary on the sliding manifold.

$$B(t) = \{x, |s(x,t)| \leq \varphi\} \quad \varphi > 0 \quad (27)$$

Outside the boundary layer, control law is chosen (following sliding condition) guaranteeing attractiveness. All trajectories starting inside the boundary will remain inside, by using the usual feedback control, B(t) i.e. replacing the term $sign(s)$ by $s/\varphi$ inside B(t). This leads to tracking within a guaranteed precision, inside boundary layer width. Saturation function 'sat(s)' is adopted instead of sign(s).

$$sat(s) = \begin{cases} 1 & s > \Delta \\ ks & |s| \leq \Delta, k = 1/\Delta \\ -1 & s < -\Delta \end{cases} \quad (28)$$

Substituting the above equation in (23) will result in:

$$u(t) = -(K_d.g(\theta,t))^{-1}.(K_i(\theta_r(t) - \theta(t)) + K_p(\dot{\theta}_r(t) - \dot{\theta}(t)) + K_d(\ddot{\theta}_r(t) - f(\theta,\dot{\theta},t)) + ks + k_{sc}|s|^{\alpha}.sat(s)) \quad (29)$$

## V. STABILITY

Lyapunov stability criteria's origin dates back to 1970's by Leitmannn and Gutman [42, 66-68]. It is the most popular approach to evaluate and to prove the stable convergence property of the sliding mode controller. Second Lyapunov approach is used to prove the stability.

- Candidate Lyapunov function for equivalent approach:

$$V(t) = \frac{1}{2}s^2(t) \text{ with V(0)=0 and V(t)>0 for s(t)} \neq 0. \quad (30)$$

A sufficient condition or reaching condition is to select the control law that forces the trajectory of error to the sliding phase from reaching phase, given as

$$\dot{V}(s) = -s^T diag\{sign(s_i)\}|\dot{s}| = -|s^T||\dot{s}| < 0 \quad (31)$$

Substituting (16) in (21), we get

$$\dot{V}(t) = s(t)(-k_{sc}\text{sign}(s(t)) - d(t)) \tag{32}$$

which is derived to prove the Lyapunov criteria.

$$\dot{V}(t) = -k_{sc}|s(t)| - s(t)d(t) \tag{33}$$

$$\dot{V}(t) \leq -k_{sc}|s(t)| + s(t)d_{max} \tag{34}$$

$$\dot{V}(t) \leq -|s(t)|(k_{sc} - d_{max}) \tag{35}$$

$$\dot{V}(t) \leq 0. \tag{36}$$

In the reaching phase with $s(t) \neq 0$, we get $k_{sc} > d_{max}$ and $|s(t)| > 0$ which lead to $\dot{V}(t) \leq 0$ i.e. a negative definite condition, satisfying the direct Lyapunov stability criteria.

- Candidate Lyapunov function for the proposed approach

$$V(t) = {}^1\!/_2\, s^2 => \dot{V}(t) = s\dot{s} \tag{37}$$

Substituting from (26),

$$\dot{V}(t) = s|-ks - k_{sc}|s|^\alpha . sat(s)| \tag{38}$$

$$\dot{V}(t) = -ks^2 - k_{sc}|s|^\beta. \tag{39}$$

So $\dot{V}(t) < 0$ because $> 0, k_{sc} > 0$ , verifying the condition that the PID sliding mode surface exist and can reach under the control law (23) for system (4).

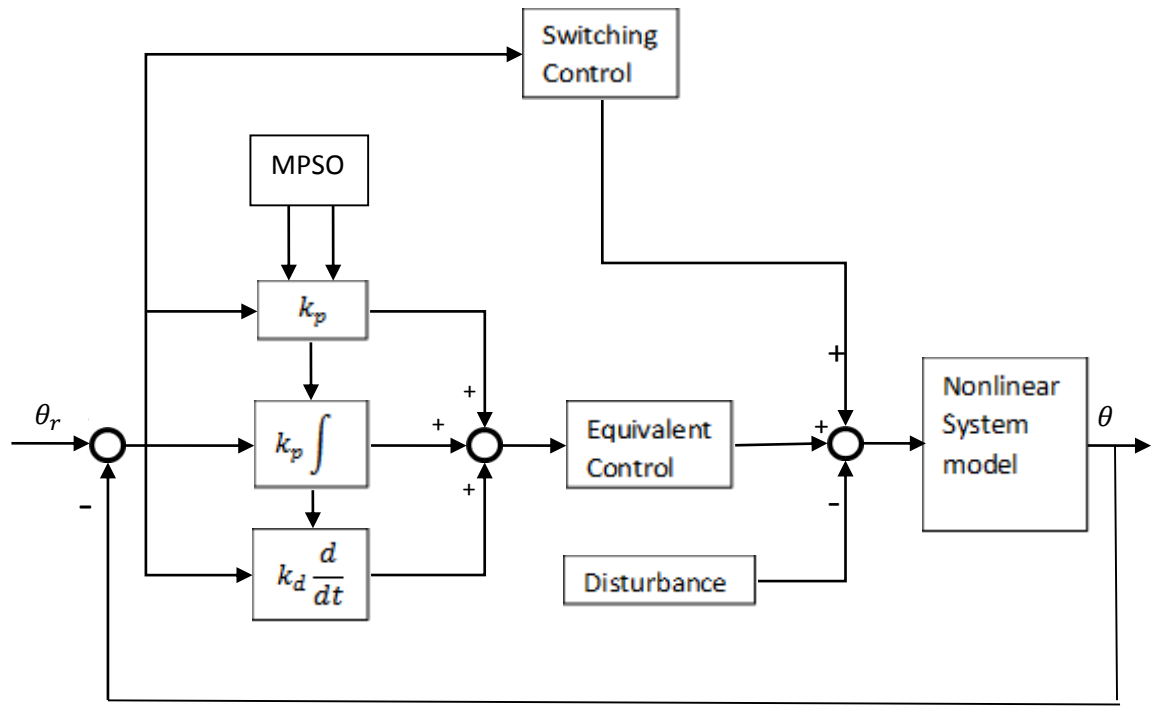

**Figure 1.** Description of the nonlinear system based on sliding mode control.

## VI. SIMULATION & EXPERIMENTAL RESULTS

The Modified PSO algorithm is applied find the parameters to the above system when the system is initially at stable equilibrium position i.e. vertically downward position. These parameters are obtained offline for the proposed PID sliding surface based sliding mode controller using improved exponential power rate reaching law (50, 60) shown in Figure 1. In the simulations, the variables used in the modified PSO algorithm are given by Parameters of sliding modes, which were held constant during experiments. PID Sliding mode control is of general form given by $u(t)$ Control force, which is defined in terms of the error between the desired and actual output and $K_d, K_i, K_p, k$ and $k_{sc}$ are switching parameters. PID and switching parameters are chosen such that the proportional gain $K_p$ has positive values to improve the activity, and the derivative gain $K_d$ limited to suppress the high frequency and noise, and value of $k$ should be greater than $k_{sc}$ to reduce the effect of chattering Setting used in Matlab implementation of MPSO algorithm are: Number of Particles, n=50, Number of subpopulation=5, Maximal number of iterations=90 and the search interval for each particle is given by [-5 5].

### 1. INVERTED PENDULUM SYSTEM

The system shown in Figure 1 consists of an inverted pendulum mounted to a motorized cart. The main aim is to balance the pendulum upwards to a position desirable for the experiment and to maintain the position without letting it fall. The cart is driven by the dc motor, controlled by a controller. A disturbance force is applied on the top of the position.

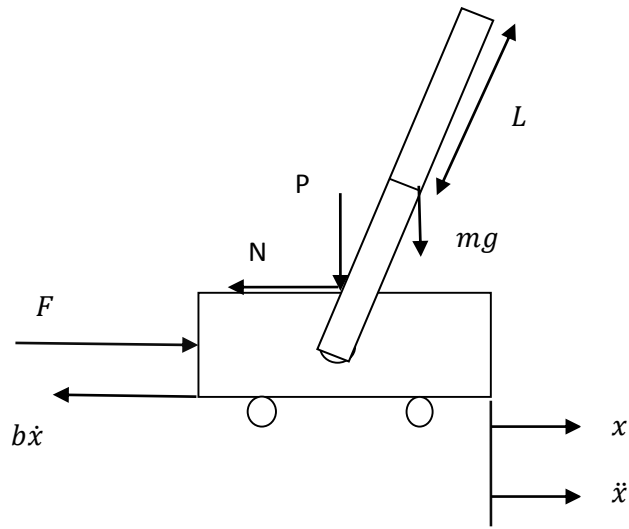

**Figure 2**. Inverted Pendulum with free body diagram

Mathematical modeling for Figure 1 has been developed using free bode diagram, to obtain the dynamic motion equation. As seen from the Figure 1, there is no vertical motion so the equations of dynamic motion can be obtained by equating the sum of forces acting along the horizontal direction.

$$M\ddot{x} + b\dot{x} + N = F \tag{40}$$

$$\mathcal{T} = r\mathcal{F} = \mathbb{I}\ddot{\theta} \tag{41}$$

Horizontal component of the force is $ml\ddot{\theta}\cos(\theta)$. (42)

Component along the direction of N= $ml\dot{\theta}^2\sin(\theta)$ (43)

Horizontal component is given as

$$N = m\ddot{x} + ml\ddot{\theta}\cos(\theta) - ml\dot{\theta}^2\sin(\theta) \tag{44}$$

Substituting (44) in (33)

$$(M + m)\ddot{x} + b\dot{x} + ml\ddot{\theta}\cos(\theta) - ml\dot{\theta}^2\sin(\theta) = u \tag{45}$$

Vertical component on substitution is given as

$$P\sin(\theta) + N\cos(\theta) - mg\sin(\theta) = ml\ddot{\theta} + m\ddot{x}\cos(\theta) \tag{46}$$

Substituting (46) in (40)

$$(I + ml^2).\theta + mgl\sin(\theta) = -ml\ddot{x}\cos(\theta). \tag{47}$$

Therefore, the dynamic equation for the single stage inverted pendulum considering b=0 is

$$\ddot{\theta} = \frac{mgl\sin(\theta) - m^2l^2\cos(\theta)\sin(\theta)\dot{\theta}^2 + u.ml\cos(\theta)}{m^2l^2\cos^2(\theta) - (I+ml^2)} \tag{48}$$

Here, $\theta = constant = 0$.

Substituting (48) and (4) in equation (23), we get

$$u_{eq} = -\frac{1}{K_d g(\theta,t)}\left(K_i(\theta_r(t) - \theta(t)) + K_p\left(-\dot{\theta}(t)\right) + K_d\left(-f(\theta,\dot{\theta},t)\right) - sgn(s(x))\right) \tag{49}$$

$$u(t) == -(K_d.g(\theta,t))^{-1}.(K_i(\theta_r(t) - \theta(t)) + K_p(\dot{\theta}_r(t)) + K_d(\ddot{\theta}_r(t) - f(\theta,\dot{\theta},t)) + ks +$$

$$k_{sc}|s|^\alpha.sat(s)) \tag{50}$$

**Table 1.**
**Operating Parameters of the Inverted Pendulum system [3]**

| S. No. | Parameters | Description of the Quantity | Value with units |
|---|---|---|---|
| 1 | Mc | Mass of the cart | 1 kg |
| 2 | Mp | Mass of the pendulum | 0.1 kg |
| 3 | I | Moment of Inertia | 0.006 kgm$^2$ |
| 4 | L | Length of the pendulum | 0.3 m |
| 5 | G | Acceleration due to gravity | 9.8 ms$^{-2}$ |
| 6 | B | Frictional coefficient | 0 |

The Inverted Pendulum system (1) having the parameters as€ given in Table 1 is considered for swing up and stabilization. Simulations results are conducted to provide the effectiveness of the proposed algorithm. The MPSO PID sliding surface based sliding mode controller with improved reaching law (50) is applied to the above system when the pendulum is initially at stable equilibrium point, i.e. at the vertical downward position making it the initial condition. The values of the parameters are chosen as $K_d = 0.8, K_i = 4, K_p = 105, k = 35 \text{ and } k_{sc} = 1.5$, and the desired position is at vertically upwards position i.e. $\theta = 0$. External disturbance applied to the IP system is $d = 10\sin(t)$.

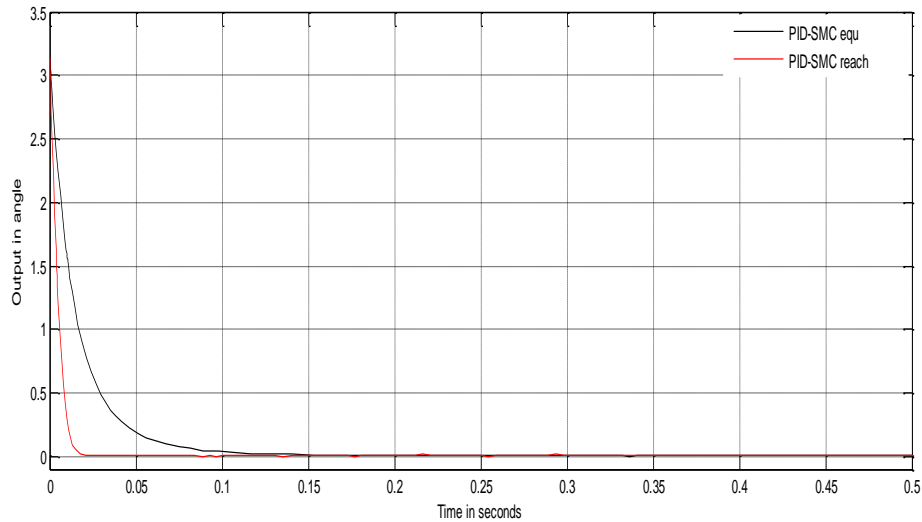

**Figure 3**. Simulation result: Output with uncertainty 'd' for MPSO PID-SMC with improved reaching law compared with equivalent control method [39].

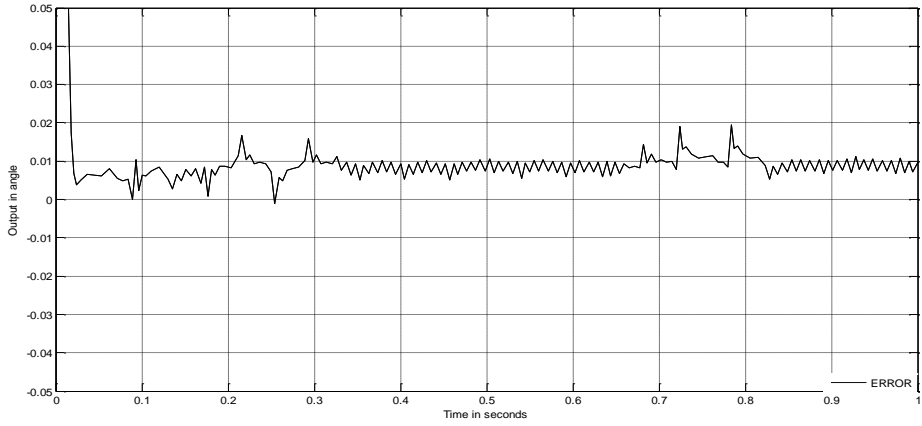

**Figure 4**. Error with uncertainty 'd' for MPSO PID-SMC with improved reaching law.

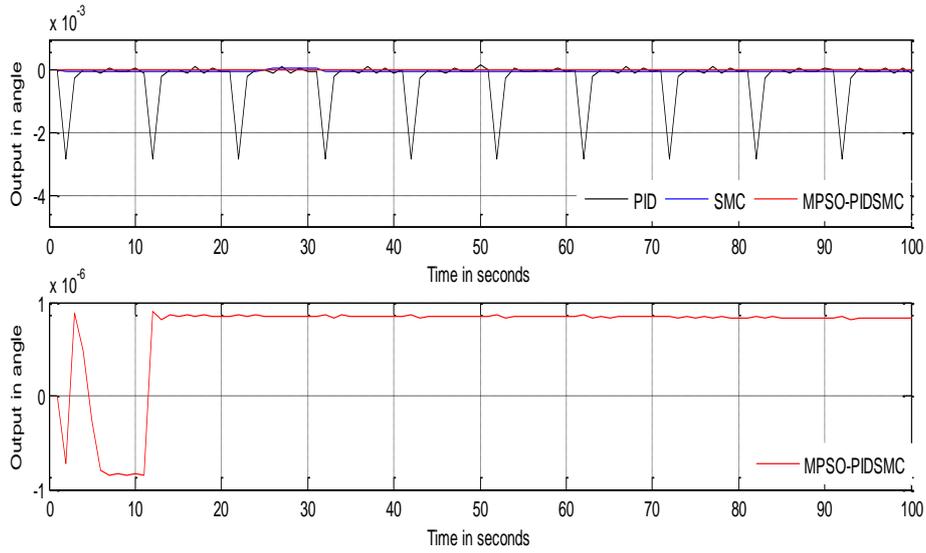

**Figure 5**. Simulations result: (a) Output comparison for PID (3), SMC (84, 52) and proposed MPSO PID-SMC with uncertainty as external disturbance, d= $1000\delta(10t)$. (b) Output of MPSO PID SMC with improved reaching law.

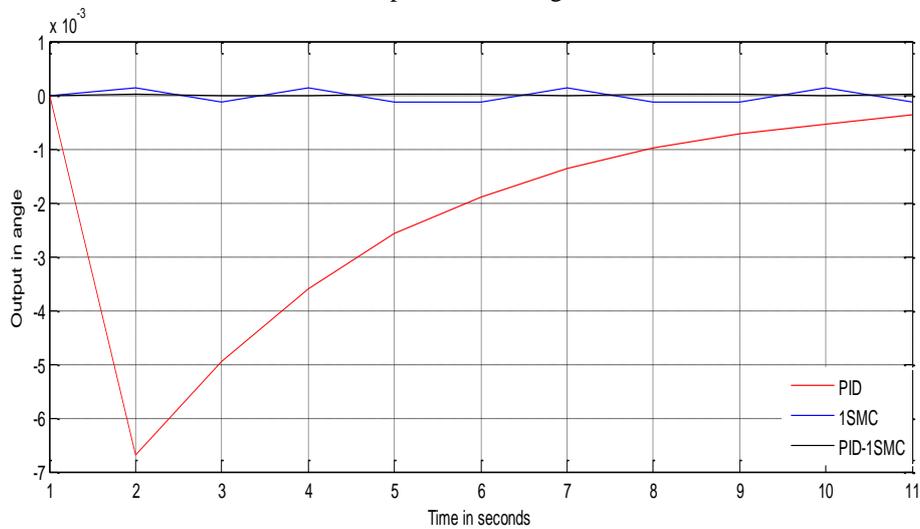

**Figure 6**. Simulations result: Transient output comparison for PID (3), SMC (84, 52) and proposed MPSO PID-SMC with uncertainty as external disturbance, d= $1000\delta(10t)$.

In Figure 3, performance of the proposed algorithm with the equivalent control based approach (49-50) is compared for the second order Inverted Pendulum, when an external disturbance defined by d= $10\sin(t)$ is applied. The sliding mode control uses the property of order reduction where the order of the system is reduces by one when the system lies on the sliding surface. Integrator part of the sliding surface forces the system to converge to the error surface at fast pace. The comparison clearly shows increase in the rate of convergence for the proposed algorithm. Figure 4 shows that the steady state error, which is very near to the zero error surface limiting itself to the boundary upon which the sliding mode's chattering phenomenon is forced. Fig 5 shows the performance comparison between controllers PID [3], SMC [84, 52] and the proposed MPSO PID sliding surface based sliding mode control with improved reaching law, which clearly depicts invariance property of SMC w.r.t. PID and better convergence of the PID-SMC w.r.t. SMC [85]. In Figure 7 proposition of faster convergence to the error surface is shown by simulation which also proves the usefulness of the Integral term in the sliding surface, as described above.

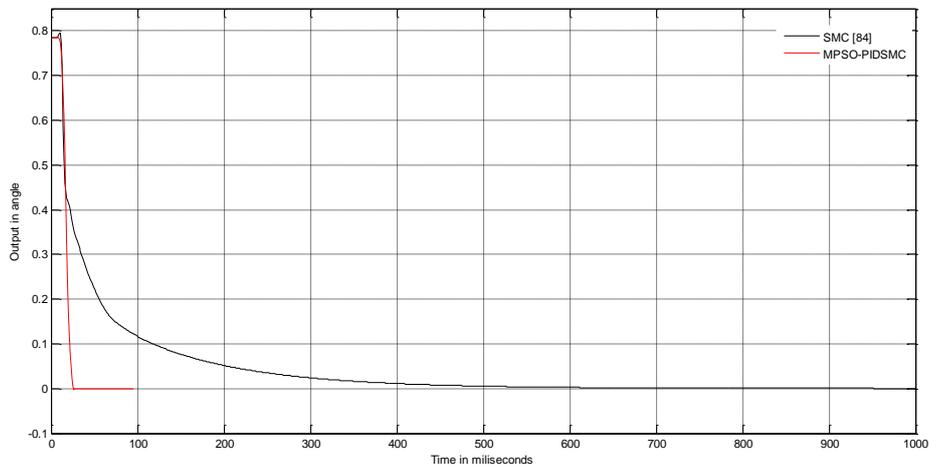

**Figure 7**. Simulations result: Transient response comparison for SMC [84, 52] and proposed MPSO PID-SMC with improved reaching law and uncertainty as external disturbance, d= $10\sin(t)$.

**Table 2**
**Comparison of Simulation results for the Inverted Pendulum system**

| S.No. | Parameters | 1-SMC | PID-SMC (eq.) | Proposed PID-SMC |
|---|---|---|---|---|
| 1 | Rise time (seconds) | 0.35 | 0.1 | 0.02 |
| 2 | Settling time (seconds) | 0.55 | 0.125 | 0.02 |

2. **CONICAL TANK SYSTEM**

Conical tank model is a benchmark problem in nonlinear control systems. In this level process, conical shape of the tank with constant changing area of cross section makes the system nonlinear. The objective is to maintain the level of the liquid at a desired level, which is achieved by controlling the input flow into the tank. The controlling variable is the level of the tank and the manipulated variable is the inflow to the tank. The liquid level will flow into the tank through inlet and the liquid will come out of the tank through outlet.

Feedback control system is designed based on the actuating error signal, which is the difference between the feedback signal and the set point, is fed to the controller to bring the output to the desired value. Error is fed to the controller for the proposed PID sliding surface design, converging the error to zero in finite time. The proposed sliding surface design will force the trajectory to converge to faster than the standard sliding mode, as will be discussed in simulations section.

The structure of the conical tank system is shown in Fig. 8. The tank level process to be simulated is a single input single output (SISO) tank system. The user can control the inflow rate by adjusting the control signal, Fin. During the simulations, the level 'h' will be calculated at any instant of time, t.

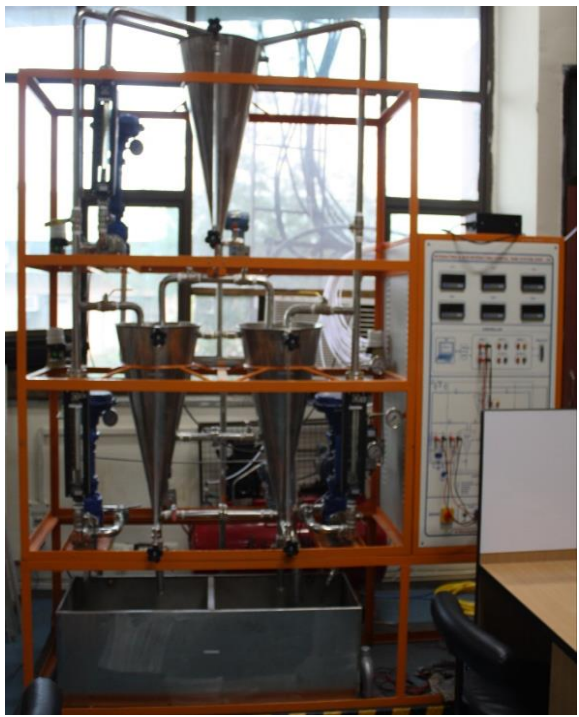 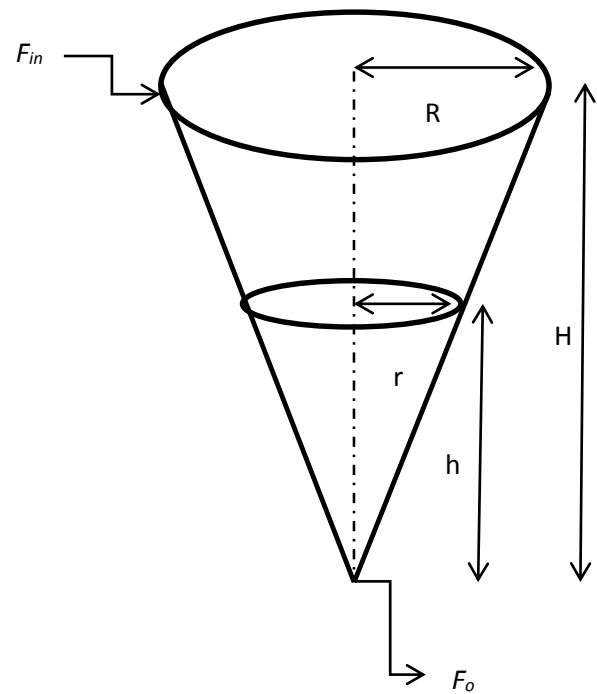

Figure 8. Model of a conical tank system

Dynamic model of conical tank is given by as area of the conical tank:

$$A = \pi r^2 \tag{51}$$

$$\tan\theta = r/h = R/H \tag{52}$$

$$r = R \cdot h/H \tag{53}$$

According to the law of conservation of mass,

Inflow rate – outflow rate = Accumulation. (54)

$$F_{in} - F_{out} = A\frac{dh}{dt} \tag{55}$$

$$F_{out} = k\sqrt{h} \quad \text{where, k = discharge coefficient} \tag{56}$$

Therefore,

$$F_{in} - k\sqrt{h} = A\frac{dh}{dt} \quad \text{where, } A = \pi \cdot R^2 \cdot h^2/H^2 \tag{57}$$

Rate of change of height,

$$\frac{dh}{dt} = \frac{F_{in} - k\sqrt{h}}{(\pi \cdot R^2 \cdot h^2/H^2)} \tag{58}$$

**Table 3**
**Operating Parameters for Conical tank system**

| S. No. | Parameters | Description of the Quantity | Value |
|---|---|---|---|
| 1. | R | Total radius of the cone | 17.5 cm |
| 2. | H | Maximum total height of the tank | 70 cm |
| 3. | K | Value of the coefficient | 55 $cm^2/s$ |
| 4. | Fin | Maximum inflow rate of the tank | 400 LPH |

Control effort required for tracking control of the desired level process is given as

$$u(t) = -\frac{1}{K_p g(h,t)}\left(K_i(h - h_d) + K_p f(h) - k_{sc} \cdot sgn(s(x))\right) \tag{59}$$

$$u(t) = -(K_p \cdot g(h,t))^{-1} \cdot \left(K_i(h(t) - h_d(t)) + K_p(f(h)) + ks + k_{sc}|s|^\alpha \cdot sat(s)\right) \tag{60}$$

The schematic block diagram is shown in Figure 8 with the parameters specified in Table 2. The system consists of a conical water tank, a water pump mechanism, a liquid level sensor and a PC-base controller. Dynamic equation of the conical tank is defined in (58) with the control effort given by [60].

The MPSO PID-SMC is proposed to be applied to the liquid level controller of a conical tank system. The objective is to control the liquid level of the tank by introducing a leakage (external disturbance) in the tank. The objective consists in minimizing the liquid level tracking error in presence of model uncertainties and leakage in the tank, limited only by the sensors capability to detect the level. The parameters for the proposed MPSO PID-SMC are chosen as $K_d = 0.8, K_i = 4.2, K_p = 105, k = 35$ and $k_{sc} = 1.5$.

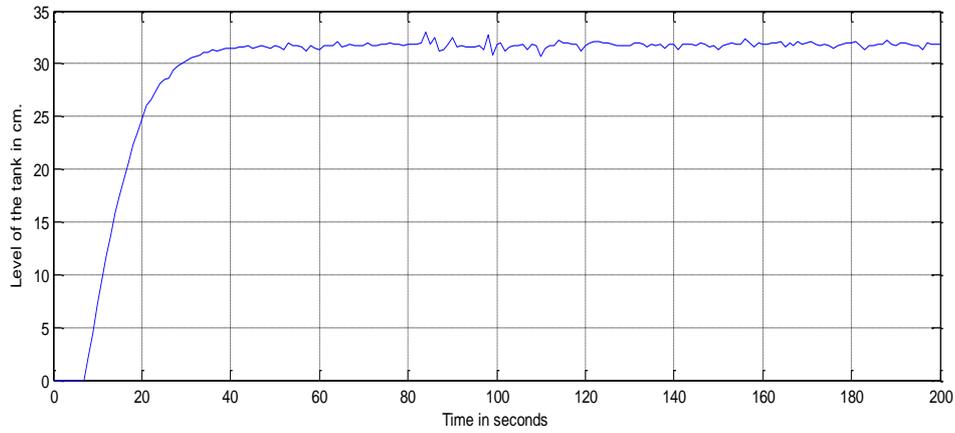

**Figure 9**. Output with uncertainty 'd' for MPSO PID-SMC with improved reaching law.

### 3. VAN DER POL EQUATION

Van der pol's equation [65] is used as the last example, for simulation and to verify the proposed algorithm.

$$\dot{x}_1 = x_2 \tag{61}$$

$$\dot{x}_2 = -2x_1 + 3(1 - x_1^2)x_2 + u \tag{62}$$

$$y = x_1 \tag{63}$$

$$u(t) = -(K_d \cdot g(\theta, \dot{\theta}, t))^{-1} \cdot (K_i(\theta_r(t) - \theta(t)) + K_p(\dot{\theta}_r(t) - \dot{\theta}(t)) + K_d(\ddot{\theta}_r(t) - f(\theta, \dot{\theta}, t)) + ks + k_{sc}|s|^\alpha \cdot sat(s)) \tag{64}$$

Considering the system (61, 62) where the external disturbance $d = 10 \sin t$. Following are the extensive simulations to demonstrate the effectiveness of the proposed PID sliding mode tracking control. To proceed the design of PID sliding mode control, parameters chosen as $K_d = 0.8, K_i = 8, K_p = 105, k = 35$ and $k_{sc} = 1.5$. The initial conditions are chosen as [pi/60 0], and the desired trajectory is chosen as $y_d = 0.1 \sin t$.

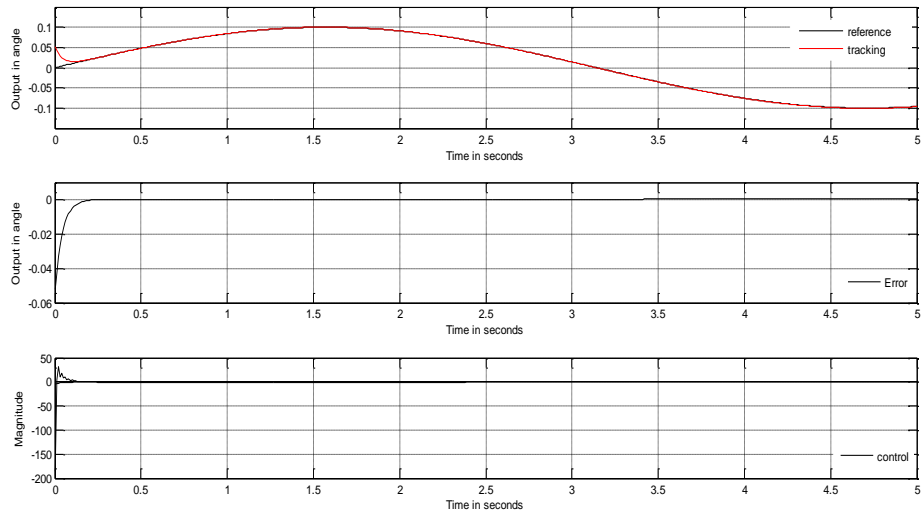

**Figure 10**. Simulatios results: with disturbance 'd'. (a)Tracking control for the van der pol equation using the PID sliding mode control with $y_d = 0.1 \sin t$. (b) error output. (c) control force.

In the following simulations, matlab\simulink 13 is utilized to implement the above MPSO PID-SMC based improved reaching law algorithm. The sampling time is set to 0.01 seconds for simulating the nonlinear differential equations. The comparison of MPSO PIDSMC with the SMC is shown in Figure 6. Figure 10(a) shows excellent tracking performance for the given system with a very less convergence time and complete tracking even in the presence of applied disturbace, shown in Figure 10. Figure 10(b) stamped the concept of finite convergence and tracking phenomenon. Figure 10(c) shows the required control effort applied to the system.

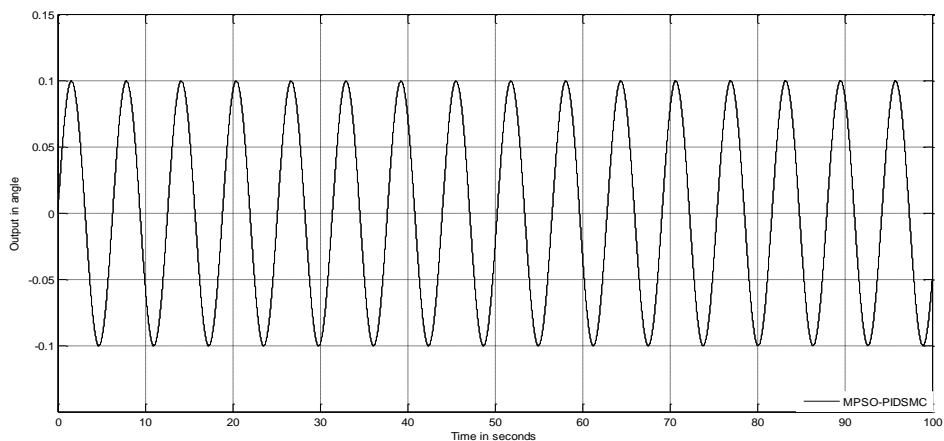

**Figure 11**.Tracking control for the van der pol equation using derived PID switching parameters.

## VII. CONCLUSION

In this paper Modified Particle Swarm Optimization (MPSO) PID sliding surface based Sliding mode control with exponential power rate reaching law is proposed for controlling non-linear systems. The problem of swing up and stabilization of the Inverted Pendulum, which is a benchmark example for control of nonlinear systems is considered for the application of the proposed algorithm. Simulations studies confirms the improved performance of the proposed PID-SMC ensuring invariance property in the presence of matched uncertainties and external disturbances, compared to the first order sliding mode and PID controllers, without sacrificing the tracking accuracy. The closed loop system has been shown to be stable in the sense of direct Lyapunov's approach. In the experimental example of conical tank, this algorithm has been implemented showing the effectiveness of the proposed control method. In order to avoid chattering phenomenon, saturation function (boundary condition) has been adopted to smoothen the switching signal. The PID-SMC guarantees better transient performance and in fact produces faster system response. The steady state error is smaller than the SMC found its suitability for tracking purposes, also better behavior of the output in case of bounded external disturbance. Experimental results confirm the fact that the sliding mode controllers are reasonable candidate to be used in industrial applications as they simple to use and easy to implement.